%% file: main.tex
\documentclass{article}
\usepackage{spconf}

\usepackage{amsmath,amssymb,amsfonts}
\usepackage{algorithmic}
\usepackage{graphicx}
\usepackage{hyperref}
\usepackage{textcomp}
\usepackage{xcolor}
\usepackage{booktabs}
\usepackage{array}
\usepackage{kotex}
\usepackage{subcaption}
\usepackage{caption}
\usepackage{multirow}
\begin{document}

\title{P2VA: Converting Persona Descriptions into Voice Attributes \\for Fair and Controllable Text-to-Speech
}

\name{Yejin Lee\textsuperscript{1}, Jaehoon Kang\textsuperscript{1}, Kyuhong Shim\textsuperscript{1,2}}
\address{
\textsuperscript{1}Department of Artificial Intelligence, Sungkyunkwan University, Korea \\
\textsuperscript{2}Department of Computer Science and Engineering, Sungkyunkwan University, Korea \\
\small{\texttt{\{yj.lee, morateng, khshim\}@skku.edu}}
}

\maketitle

\begin{abstract}
\input{section/00_abstract}
\end{abstract}


\input{section/01_introduction}

\input{section/02_related}
\input{section/03_method}
\input{section/04_experiment}
\input{section/05_conclusion}

\newpage
\bibliographystyle{IEEEbib}
\bibliography{reference}

\end{document}

%% file: section/00_abstract.tex
While persona-driven large language models (LLMs) and prompt-based text-to-speech (TTS) systems have advanced significantly, a usability gap arises when users attempt to generate voices matching their desired personas from implicit descriptions. Most users lack specialized knowledge to specify detailed voice attributes, which often leads TTS systems to misinterpret their expectations. To address these gaps, we introduce \textbf{Persona-to-Voice-Attribute} (P2VA), the first framework enabling voice generation automatically from persona descriptions. Our approach employs two strategies: P2VA-C for structured voice attributes, and P2VA-O for richer style descriptions. Evaluation shows our P2VA-C reduces WER by 5\% and improves MOS by 0.33 points. To the best of our knowledge, \textbf{P2VA} is the first framework to establish a connection between persona and voice synthesis. 
In addition, we discover that current LLMs embed societal biases in voice attributes during the conversion process. 
Our experiments and findings further provide insights into the challenges of building persona-voice systems.  

%% file: section/01_introduction.tex
\section{Introduction}\label{sec:intro}
Recent advances in large language models (LLMs) have accelerated the development of interactive AI applications such as personalized chatbots, metaverse dialogue systems, and virtual influencers. A central element in these systems is the persona, a virtual identity that encodes traits such as personality, background, and interests~\cite{zhang2018personalizing, tseng2024two, liu2024llms+}. For example, specifying a persona such as “a calm and trustworthy healthcare assistant” or “a concise and professional financial agent” can greatly enhance coherence and realism in LLM-driven dialogue~\cite{ge2024scaling, hong2024metagpt, park2023generative}.
In parallel, text-to-speech (TTS) technology has also achieved remarkable progress. Prompt-based TTS provides fine-grained control of prosodic attributes, such as gender, accent, tone, pitch, and speed, through natural language prompts~\cite{guo2023prompttts, shimizu2024prompttts++, lyth2024natural, yang2024instructtts}. 

 \begin{figure}[t!]
     \centering
     \includegraphics[width=1.0\linewidth]{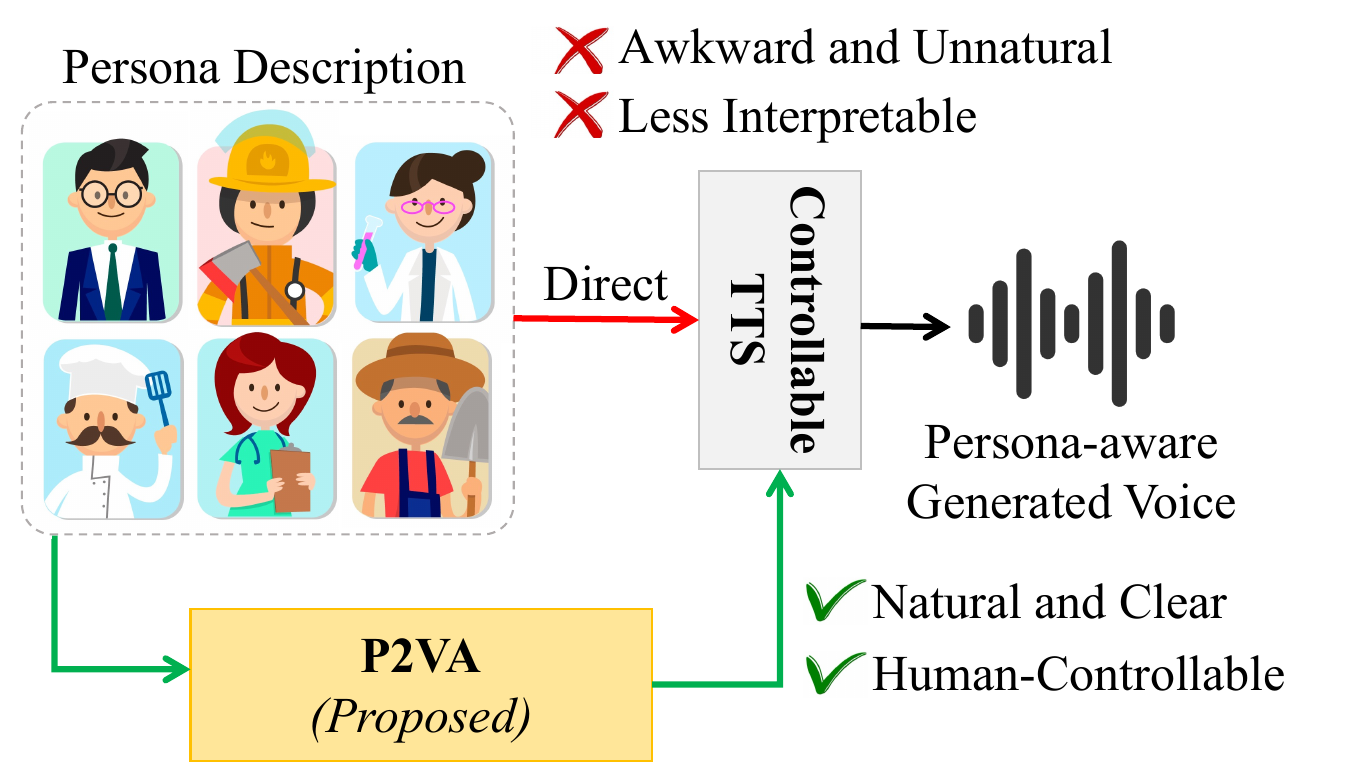}
     \caption{Illustration of the Persona-to-Voice-Attribute (P2VA). Persona descriptions designed for text-based LLMs are suboptimal for controllable TTS models.
  }
     \label{fig:intro}
     \vspace{-0.3cm}
 \end{figure}

As persona modeling and prompt-based TTS capabilities advance, there is a growing demand to associate personas with matching voices.
When generating voices for a persona using TTS models, highly experienced users can explicitly decide detailed voice style attributes. In contrast, most users tend to rely on implicit and abstract descriptions. 
TTS systems often fail to interpret users' expectations when such ambiguous descriptions are given, leading to a usability gap.  

In this paper, we propose \textbf{Persona-to-Voice-Attribute} (P2VA) to bridge this gap, enabling voice generation based on persona descriptions without requiring specialized knowledge (see Fig.~\ref{fig:intro}).
We introduce two complementary P2VA strategies: P2VA-C, which maps personas to structured attributes; and P2VA-O, which produces richer free-form style descriptions (see Fig.~\ref{fig:proposed_method}). 
In addition, we discover that the lack of explicit voice attributes in persona descriptions leads LLMs to impose strong biases, especially related to gender-specific patterns. To the best of our knowledge, \textbf{P2VA} is the first work to establish a connection between persona and voice style.

\vspace{0.2cm}
We summarize our contributions as follows:
\vspace{-0.2cm}
\begin{itemize}
    \item We introduce P2VA, a novel framework for converting natural language persona descriptions into explicit voice attributes.
    \item We address the usability gap of speech synthesis by allowing non-professional users to generate appropriate speech properties without the need for expertise in speech engineering.    
    \item We analyze bias patterns in LLMs from multiple perspectives and present insights that should be considered in future research in the persona-voice system.
\end{itemize}

%% file: section/02_related.tex
\section{Related Work}\label{sec:related}
 
\begin{figure}[t]
    \centering
    \includegraphics[width=0.95\linewidth]{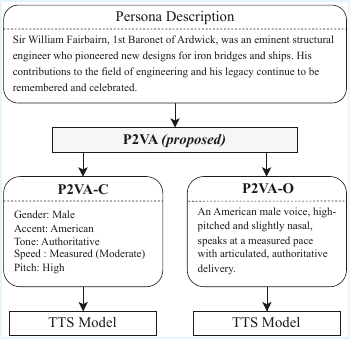}
    \caption{Process of Persona-to-Voice-Attribute (P2VA). P2VA-C (left) converts traits of persona descriptions into a voice attribute (e.g., gender, speed, accent), while P2VA-O (right) produces free-form, natural language style descriptions.}
    \label{fig:proposed_method}
\end{figure}

\subsection{Prompt-based TTS}\label{ssec:related_prompt}
Prompt-based TTS offers fine-grained control of speech generation through natural language instructions. PromptTTS~\cite{guo2023prompttts} showed that text descriptions alone can drive prosodic variation without reference audio. PromptTTS++~\cite{shimizu2024prompttts++} added control of speaker identity expressed in natural language, turning style and identity into a joint control problem. 
Parler-TTS~\cite{lyth2024natural} used synthetic annotations to achieve high-fidelity speech from textual instructions. InstructTTS~\cite{yang2024instructtts} modeled expressive speech in a discrete latent space guided by style prompts. PL-TTS~\cite{li2024pltts} combined a diffusion model with LLMs and improved generalization under prompt conditioning. 
Recently, MM-TTS~\cite{guan2024mm} unified text, speech, and images within a single style space and applied multimodal prompts for style transfer. 

However, these approaches rest on a shared assumption that users already specify concrete style terms such as emotion, speed, tone, and speaker traits. In practice, many users describe voices in terms of personality and character. A procedure for converting descriptions into controllable vocal attributes has not yet gained attention. 

\subsection{Persona-based Personalization in LLMs}\label{ssec:related_persona}
Persona is a virtual identity that maintains consistency in tone, knowledge, and attitude throughout a conversation, emerging as a core component of LLMs.
PersonaChat~\cite{zhang2018personalizing} defined speaker identity with short self-descriptive sentences and produced personalized dialogue. Personalized LLMs~\cite{liu2024llms+} attach a compact user-history embedding to prompts to personalize LLMs without finetuning. PersonaHub~\cite{ge2024scaling} uses a 1B-persona corpus to generate diverse synthetic data that scales across specific tasks. Persona-based multi-agent systems are also growing. MetaGPT~\cite{hong2024metagpt} utilizes role-specialized personas through standardized procedures, structuring collaboration on complex tasks. Generative agents~\cite{park2023generative} instantiate persistent personas, yielding consistent individual behavior and emergent social dynamics. 
These developments show that persona is not merely a character setting but a core axis of interaction design. 

\begin{table*}[t!]
\centering
\caption{Examples of persona-to-voice attribute (P2VA) conversion: original persona descriptions (left) are transformed into structured vocal attributes (center) and voice style descriptions for TTS models (right).}
\vspace{-0.1cm}
\label{tab:table1}
\renewcommand{\arraystretch}{1.4}
\scriptsize
\begin{tabular}{p{0.33\linewidth} p{0.35\linewidth} p{0.25\linewidth}}
\toprule[1.2pt]
\textbf{Textual Persona (Input)} & \textbf{Rewritten Attributes} & \textbf{Voice Style Description (Output)} \\
\midrule[1.2pt]
A seasoned financial professional with a deep understanding of the Dutch financial market. 
Mrs. Simone Huis Veld has over fifteen years of senior management experience in the Dutch financial sector. 
Her expertise and leadership skills make her a valuable asset to Euronext and the European financial market.
&
\scriptsize
\textbf{Gender:} Female (Mrs. Simone) \;|\;
\textbf{Accent:} British (European professional) \;|\;
\textbf{Pitch:} Medium (balanced authority) \;|\;
\textbf{Speed:} Measured (senior management) \;|\;
\textbf{Tone:} Authoritative, calm (CEO, leadership) \;|\;
\textbf{Prosody:} Punctuated, animated (leadership skills) \;|\;
\textbf{Timbre:} Deep, silky (seasoned professional)
&
A British-accented female voice, medium-pitched and measured, with a deep, silky resonance. 
Crisp and authoritative, her punctuated, animated delivery maintains a calm steadiness.
\\
\midrule

An Indian astrophysicist and theoretical physicist. Abhas Mitra is known for his black hole theory and work at Bhabha Atomic Research Centre in Mumbai. His expertise and ability to challenge conventional beliefs have made him a respected figure in astrophysics.
& 
\scriptsize
\textbf{Gender:} Male (Abhas) \;|\;
\textbf{Accent:} Indian (Mumbai) \;|\;
\textbf{Pitch:} Medium (scientific authority) \;|\;
\textbf{Speed:} Measured, flowing (theoretical physicist) \;|\;
\textbf{Tone:} Authoritative, calm (respected figure) \;|\;
\textbf{Prosody:} Subtly animated (challenge beliefs) \;|\;
\textbf{Timbre:} Deep, crisp (theoretical precision)
& 
An Indian male voice with a medium pitch, deep and silky resonance, and crisp articulation. He speaks at a measured, flowing pace with an authoritative, calm tone and subtly animated delivery.
\\
\bottomrule[1.2pt]
\end{tabular}
\end{table*}

%% file: section/03_method.tex
\section{Persona-to-Voice-Attribute (P2VA)}\label{sec:method} 
\subsection{Motivation} 
Persona description encompasses character background, linguistic style, and personality traits. On the other hand, voice attributes determine prosodic features such as intonation, speech rate, emotional expression, tone, and gender. These dimensions are conceptually interconnected; however, to the best of our knowledge, limited research has directly established the connection between these two domains.

\subsection{P2VA-C: Closed-ended Form} 
The P2VA-C strategy guides the LLM to extract information based on a predefined set of vocal attributes. Given a persona description, the LLM infers values for each attribute slot and outputs them in a structured JSON format. The structured data is then transformed into a vocal style description prompt suitable for TTS input (see Table~\ref{tab:table1} for examples). P2VA-C offers strong controllability, making it well-suited for applications requiring predictable and interpretable outputs.

\subsection{P2VA-O: Open-ended Form}
Conversely, in the P2VA-O strategy, the LLM generates voice style descriptions without constraints from a predefined set (see Fig.~\ref{fig:proposed_method}). 
This relaxation enables the model to extract a variety of voice attributes that could not be fully expressed with a combination of predefined voice attribute sets. The flexibility of P2VA-O makes it well-suited for emotionally immersive interfaces and narrative-driven applications, such as storytelling and conversational agents.

%% file: section/04_experiment.tex
\section{Experimental Results}\label{sec:experiment} 

\subsection{Setup}\label{ssec:experiment_setup}
We utilize Parler-TTS~\cite{lyth2024natural}, a natural language-controllable TTS model that provides a flexible interface for fine-grained style manipulation without reference audio. To transform persona descriptions into voice style prompts, we employ \texttt{GPT-4o-mini} as the underlying LLM. The persona descriptions are sampled from the Persona-1M dataset~\cite{ge2024scaling}, a large-scale synthetic corpus. We leverage transcripts from LJ-Speech, a widely adopted benchmark dataset used in the TTS domain.

\subsection{Style Preset}\label{ssec:experiment_style}
For P2VA-C, we define a fixed list of style attributes by examining the most frequent labels in Parler-TTS’s training data.
Our preset includes options for gender, accent (e.g., American, British, Asian, Indian), tone (e.g., Analytical, Warm, Engaging), speed (e.g., Slow, Normal, Fast), and pitch (e.g., Low, Medium, High). 

\subsection{Quantitative Evaluation}\label{ssec:experiment_metric}
We measure synthesized speech quality and accuracy using Word Error Rate (WER), UTMOS, and Mean Opinion Score (MOS) estimated by LLM and human listeners. 
These metrics evaluate synthesized speech from multiple perspectives, such as clarity, naturalness, and alignment with user intent. 

\setlength{\tabcolsep}{6pt}
\begin{table}[t!]
    \centering
    \caption{Evaluation results across four metrics.}
    \vspace{-0.1cm}
    \resizebox{\linewidth}{!}{
    \begin{tabular}{l|cccc}
        \toprule[1.2pt]
        \multirow{2}{*}{Method} & WER(\(\downarrow\))  & MOS(\(\uparrow\))     & MOS(\(\uparrow\))  & UTMOS(\(\uparrow\))  \\
            &    & (Gemini) & (Human) & \\
        \midrule
        Baseline& 22\% & 4.20 & 3.09  & 2.85   \\
        \midrule
        P2VA-C  & {\textbf{17\%}} & {\textbf{4.50}} & {\textbf{3.42}} & {\textbf{2.94}}\\
        P2VA-O & 18\% & 4.43 & 3.23 & 2.88   \\     
        \bottomrule[1.2pt]
    \end{tabular}}
    \label{table2:res_of_experiment}
\end{table}
\setlength{\tabcolsep}{6pt}

We randomly sample a total of 1000 persona–transcript pairs from the Persona-1M dataset. Then, we generate speech using three approaches: Baseline (as-is), P2VA-C, and P2VA-O. As shown in Table~\ref{table2:res_of_experiment}, the P2VA-C achieves the best performance across all metrics. While the improvement in UTMOS may not be statistically significant, the reduction in WER is about 5\%, and MOS improves by 0.33 points.  

\subsection{Usability improvement}\label{ssec:experiment_user}
To assess the practical effectiveness of our method, we conducted a comparative user study with real-world TTS users. Participants first described their desired voice style for a given persona and synthesized speech accordingly using a TTS model. Then, they applied P2VA to the same specification, generated speech, and compared the two outputs. Over \textbf{90\%} of participants reported that the P2VA-generated speech aligned with their intended style.

\section{Analysis}\label{sec:analysis}

\begin{figure*}
    \centering
    \includegraphics[width=1.0\linewidth]{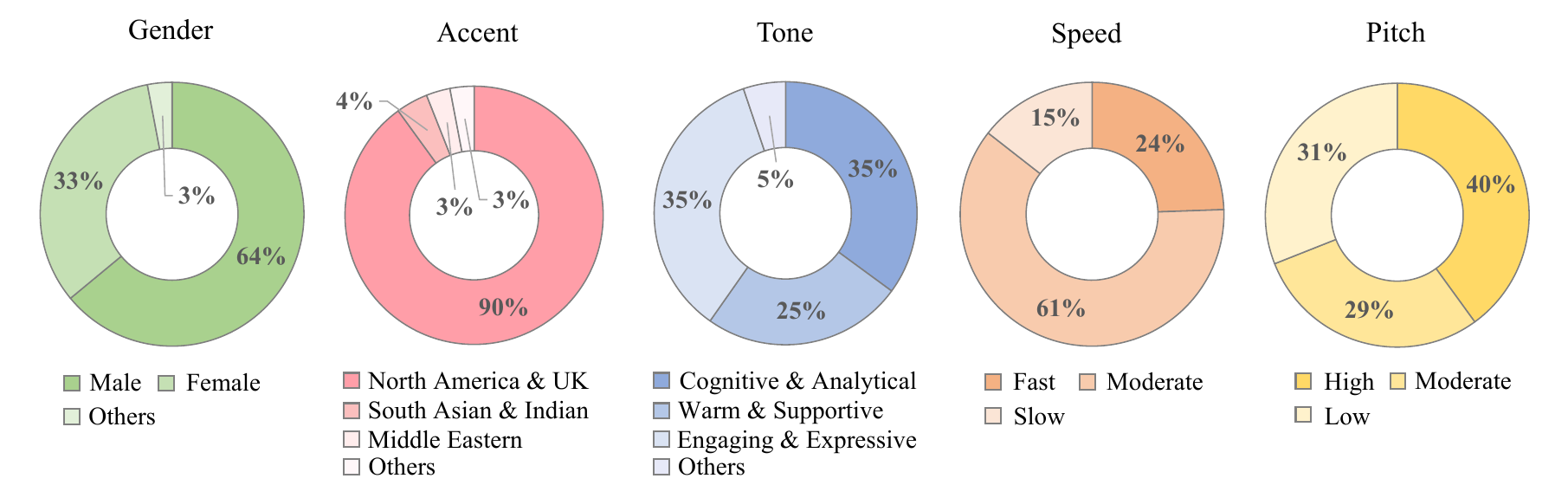}
    \vspace{-0.6cm}
    \caption{Distribution of speech style attributes generated via LLM-based prompt rewriting. The attributes are categorized into five dimensions: gender, accent, tone, speed, and pitch.
    }
    \label{fig:distribution}
\end{figure*}

\subsection{Bias in Style Attributes}\label{ssec:experiment_bias}
We analyze potential biases in LLM-based prompt-to-style conversion by classifying outputs of the P2VA-C across five dimensions: gender, accent, tone, speed, and pitch (see Fig.~\ref{fig:distribution}). 
Regardless of explicit gender cues in the input persona, the LLM assigns male voices 64\% of the time, female voices 33\%, and leaves 3\% unspecified (see Table~\ref{table2:gender_change}).
For accent, 90\% of outputs default to North American or British variants, while other regional accents are highly underrepresented.
Additionally, 61\% of the generated prompts favor a fast speed.

\setlength{\tabcolsep}{9pt}
\begin{table}[t]
    \centering
    \caption{Gender distribution change after P2VA. `Others' include not specified and uncertain samples.}
    \vspace{-0.2cm}
    \resizebox{\linewidth}{!}{
    \begin{tabular}{c|ccc}
    \toprule[1.2pt]
      & Male & Female & Others \\
    \midrule
    Original Description & 10\%  & 9\% & 81\%  \\
    After Applying P2VA & 64\% & 33\%  & 3\% \\
    \bottomrule[1.2pt]
     \end{tabular}}
    \vspace{-0.2cm}
    \label{table2:gender_change}
    \end{table}
\setlength{\tabcolsep}{6pt}

These imbalances reveal clear biases toward male representation and Western accents, suggesting that the underlying model or data carries implicit gender and regional biases.
Consequently, the model risks misrepresenting and excluding multicultural and diverse user groups.

We further analyze interactions between these biases by conditioning style distributions on gender.
Tables~\ref{table2:res_of_dis_tone},~\ref{table2:res_of_dis_speed}, and~\ref{table2:res_of_dis_pitch} show LLM's implicit bias, some of which align with common stereotypes. 
Male voices co-appear most often with `\textit{Cognitive \& Analytical}' tone and `\textit{Low}' pitch.
On the other hand, female voices disproportionately co-occur with `\textit{Warm \& Supportive}' tone and `\textit{High}' pitch.
Each value in these tables represents the relative proportion within the given gender.

\subsection{Ethical Considerations}\label{ssec:experiment_corr}
LLM bias refers not to a mere statistical imbalance, but to the inference of attributes in the absence of explicit cues. LLMs imbued with social bias tend to overgeneralize stereotypes embedded in training data~\cite{bai2024measuring, shrawgi2024uncovering}. For example, personas associated with professions such as scientists, soldiers, and programmers often default to male-style voices, even when no gender information is provided. While humans can interpret language by drawing on context and prior knowledge, recent studies~\cite{lim2025users, liu2024bias} demonstrate that LLM biases influence user perceptions and may reinforce stereotypes over time.

\subsection{Future Research Directions}\label{ssec:future_work}
Building on our findings, future work should assess existing bias-mitigation methods~\cite{smith2022m, raza-etal-2024-mbias} within our framework.
Moreover, as voice interfaces proliferate, persona specifications should include nuanced voice-style dimensions beyond character attributes.
To support this, we advocate for the creation of comprehensive benchmarks that measure both speech-style fidelity and demographic parity across multiple dimensions. 

\setlength{\tabcolsep}{11pt}
\begin{table}[t!]
    \centering
    \caption{Speech tone profiles by gender. See Fig.~\ref{fig:distribution} `Tone' (3rd circle) for the abbreviations.}
    \vspace{-0.2cm}
    \resizebox{\linewidth}{!}{
    \begin{tabular}{c|cccc}
    \toprule[1.2pt]
    Gender & C\&A & W\&S & E\&E & Others \\
    \midrule
    Male   & \textbf{41.3}\% & 14.3\%  & 36.5\% & 7.9\% \\
    Female & 23.5\%  & \textbf{44.1}\% & 32.4\%  & 0.0\% \\
    \bottomrule[1.2pt]
    \end{tabular}}
    \vspace{-0.2cm}
    \label{table2:res_of_dis_tone}
\end{table}
\setlength{\tabcolsep}{6pt}

\setlength{\tabcolsep}{16pt}
\begin{table}[t!]
    \centering
    \caption{Speech speed profiles by gender.}
    \vspace{-0.2cm}
    \resizebox{\linewidth}{!}{
    \begin{tabular}{c|cccc}
    \toprule[1.2pt]
    Gender & Fast & Normal & Slow \\
    \midrule
    Male   & 35.1\%  & \textbf{43.3}\%  & 21.6\% \\
    Female & 22.7\%  & 36.4\%  & \textbf{40.9}\% \\
    \bottomrule[1.2pt]
    \end{tabular}}
    \vspace{-0.2cm}
    \label{table2:res_of_dis_speed}
\end{table}
\setlength{\tabcolsep}{6pt}

\setlength{\tabcolsep}{15pt}
\begin{table}[t!]
    \centering
    \caption{Speech pitch profiles by gender.}
    \vspace{-0.2cm}
    \resizebox{\linewidth}{!}{
    \begin{tabular}{c|cccc}
    \toprule[1.2pt]
    Gender  & High & Moderate & Low\\
    \midrule
    Male   & 27.0\%  & 31.7\% & \textbf{41.3}\%  \\
    Female & \textbf{61.8}\% & 23.5\%  & 14.7\% \\
    \bottomrule[1.2pt]
    \end{tabular}}
    \vspace{-0.2cm}
    \label{table2:res_of_dis_pitch}
\end{table}
\setlength{\tabcolsep}{6pt}

%% file: section/05_conclusion.tex
\section{Conclusion}\label{sec:conclusion}
In this work, we propose P2VA, a strategy that transforms persona descriptions into vocal style prompts, enabling general users to leverage TTS models more easily and effectively. Through quantitative evaluation and user studies, we demonstrate that P2VA improves both speech synthesis quality and overall usability. Furthermore, we analyze how biases and stereotypes embedded in LLMs are reflected in the conversion of personas to voice styles. As an early exploratory study in the persona–TTS domain, this work highlights the need for future research to address not only speech quality but also broader ethical considerations.

%% file: main.bbl
\begin{thebibliography}{10}

\bibitem{zhang2018personalizing}
Saizheng Zhang, Emily Dinan, Jack Urbanek, Arthur Szlam, Douwe Kiela, and Jason Weston,
\newblock ``Personalizing dialogue agents: {I} have a dog, do you have pets too?,''
\newblock in {\em Proceedings of the 56th Annual Meeting of the Association for Computational Linguistics (Volume 1: Long Papers)}, July 2018, pp. 2204--2213.

\bibitem{tseng2024two}
Yu-Min Tseng, Yu-Chao Huang, Teng-Yun Hsiao, Wei-Lin Chen, Chao-Wei Huang, Yu~Meng, and Yun-Nung Chen,
\newblock ``Two tales of persona in {LLM}s: A survey of role-playing and personalization,''
\newblock in {\em Findings of the Association for Computational Linguistics: EMNLP 2024}, Nov. 2024, pp. 16612--16631.

\bibitem{liu2024llms+}
Jiongnan Liu, Yutao Zhu, Shuting Wang, Xiaochi Wei, Erxue Min, Yu~Lu, Shuaiqiang Wang, Dawei Yin, and Zhicheng Dou,
\newblock ``Llms+ persona-plug= personalized llms,''
\newblock {\em arXiv preprint arXiv:2409.11901}, 2024.

\bibitem{ge2024scaling}
Tao Ge, Xin Chan, Xiaoyang Wang, Dian Yu, Haitao Mi, and Dong Yu,
\newblock ``Scaling synthetic data creation with 1,000,000,000 personas,''
\newblock {\em arXiv preprint arXiv:2406.20094}, 2024.

\bibitem{hong2024metagpt}
Sirui Hong, Mingchen Zhuge, Jonathan Chen, Xiawu Zheng, Yuheng Cheng, Ceyao Zhang, Jinlin Wang, Zili Wang, Steven Ka~Shing Yau, Zijuan Lin, et~al.,
\newblock ``Metagpt: Meta programming for a multi-agent collaborative framework,''
\newblock International Conference on Learning Representations, ICLR, 2024.

\bibitem{park2023generative}
Joon~Sung Park, Joseph O'Brien, Carrie~Jun Cai, Meredith~Ringel Morris, Percy Liang, and Michael~S Bernstein,
\newblock ``Generative agents: Interactive simulacra of human behavior,''
\newblock in {\em Proceedings of the 36th annual acm symposium on user interface software and technology}, 2023, pp. 1--22.

\bibitem{guo2023prompttts}
Zhifang Guo, Yichong Leng, Yihan Wu, Sheng Zhao, and Xu~Tan,
\newblock ``Prompttts: Controllable text-to-speech with text descriptions,''
\newblock in {\em ICASSP 2023-2023 IEEE International Conference on Acoustics, Speech and Signal Processing (ICASSP)}. IEEE, 2023, pp. 1--5.

\bibitem{shimizu2024prompttts++}
Reo Shimizu, Ryuichi Yamamoto, Masaya Kawamura, Yuma Shirahata, Hironori Doi, Tatsuya Komatsu, and Kentaro Tachibana,
\newblock ``Prompttts++: Controlling speaker identity in prompt-based text-to-speech using natural language descriptions,''
\newblock in {\em ICASSP 2024-2024 IEEE International Conference on Acoustics, Speech and Signal Processing (ICASSP)}. IEEE, 2024, pp. 12672--12676.

\bibitem{lyth2024natural}
Dan Lyth and Simon King,
\newblock ``Natural language guidance of high-fidelity text-to-speech with synthetic annotations,''
\newblock {\em arXiv preprint arXiv:2402.01912}, 2024.

\bibitem{yang2024instructtts}
Dongchao Yang, Songxiang Liu, Rongjie Huang, Chao Weng, and Helen Meng,
\newblock ``Instructtts: Modelling expressive tts in discrete latent space with natural language style prompt,''
\newblock {\em IEEE/ACM Transactions on Audio, Speech, and Language Processing}, 2024.

\bibitem{li2024pltts}
Shuhua Li, Qirong Mao, and Jiatong Shi,
\newblock ``Pltts: A generalizable prompt-based diffusion tts augmented by large language model,''
\newblock in {\em Proc. Interspeech}, 2024, pp. 4888--4892.

\bibitem{guan2024mm}
Wenhao Guan, Yishuang Li, Tao Li, Hukai Huang, Feng Wang, Jiayan Lin, Lingyan Huang, Lin Li, and Qingyang Hong,
\newblock ``Mm-tts: Multi-modal prompt based style transfer for expressive text-to-speech synthesis,''
\newblock in {\em Proceedings of the AAAI Conference on Artificial Intelligence}, 2024, vol.~38, pp. 18117--18125.

\bibitem{bai2024measuring}
Xuechunzi Bai, Angelina Wang, Ilia Sucholutsky, and Thomas~L. Griffiths,
\newblock ``Measuring implicit bias in explicitly unbiased large language models,''
\newblock in {\em NeurIPS 2024 Workshop on Behavioral Machine Learning}, 2024.

\bibitem{shrawgi2024uncovering}
Hari Shrawgi, Prasanjit Rath, Tushar Singhal, and Sandipan Dandapat,
\newblock ``Uncovering stereotypes in large language models: A task complexity-based approach,''
\newblock in {\em Proceedings of the 18th Conference of the European Chapter of the Association for Computational Linguistics (Volume 1: Long Papers)}, 2024, pp. 1841--1857.

\bibitem{lim2025users}
Hyunseung Lim, Dasom Choi, and Hwajung Hong,
\newblock ``How do users identify and perceive stereotypes? understanding user perspectives on stereotypical biases in large language models,''
\newblock in {\em Proceedings of the 2025 ACM Conference on Fairness, Accountability, and Transparency}, 2025, pp. 3241--3253.

\bibitem{liu2024bias}
Yiran Liu, Ke~Yang, Zehan Qi, Xiao Liu, Yang Yu, and Cheng~Xiang Zhai,
\newblock ``Bias and volatility: A statistical framework for evaluating large language model's stereotypes and the associated generation inconsistency,''
\newblock {\em Advances in Neural Information Processing Systems}, vol. 37, pp. 110131--110155, 2024.

\bibitem{smith2022m}
Eric~Michael Smith, Melissa Hall, Melanie Kambadur, Eleonora Presani, and Adina Williams,
\newblock ``" i'm sorry to hear that": Finding new biases in language models with a holistic descriptor dataset,''
\newblock {\em arXiv preprint arXiv:2205.09209}, 2022.

\bibitem{raza-etal-2024-mbias}
Shaina Raza, Ananya Raval, and Veronica Chatrath,
\newblock ``{MBIAS}: Mitigating bias in large language models while retaining context,''
\newblock in {\em Proceedings of the 14th Workshop on Computational Approaches to Subjectivity, Sentiment, {\&} Social Media Analysis}, Aug. 2024, pp. 97--111.

\end{thebibliography}
